\documentclass[preprint,showpacs,preprintnumbers,amsmath,amssymb]{revtex4}

\usepackage{graphicx}
\usepackage{dcolumn}
\usepackage{bm}

\begin{document}

\title{{\bf Study of charge-charge coupling effects on
    dipole emitter relaxation within a classical electron-ion plasma
    description}}

\author{Emmanuelle Dufour, Annette Calisti, Bernard Talin}
\affiliation{%
    UMR6633, Universit\'{e} de Provence, Centre Saint J\'{e}r\^{o}me,\\
    13397 Marseille Cedex 20, France.}

\author{Marco A Gigosos, Manuel A Gonz\'{a}lez}
\affiliation{%
    Departamento de \'{O}ptica y F\'{i}sica
    Aplicada, Facultad de Ciencias,
    Universidad de Valladolid, 47071 Valladolid, Spain}

\author{Teresa del R\'{i}o Gaztelurrutia}
\affiliation{%
    Escuela Superior de Ingenieros, Ald. Urquijo s/n
    48013 Bilbao, Spain}

\author{James W. Dufty}
\affiliation{%
    Department of Physics, University of Florida,
    Gainesville, FL 32611}

\date{\today }

\begin{abstract}
    Studies of charge-charge (ion-ion, ion-electron, and
    electron-electron) coupling properties for ion impurities in an
    electron gas and for a two component plasma are carried out on the
    basis of a regularized electron-ion potential without short-range
    Coulomb divergence. This work is motivated in part by questions
    arising from recent spectroscopic measurements revealing
    discrepancies with present theoretical descriptions. Many of the
    current radiative property models for plasmas include only single
    electron-emitter collisions and neglect some or all charge-charge
    interactions. A molecular dynamics simulation of dipole relaxation
    is proposed here to allow proper account of many electron-emitter
    interactions and all charge-charge couplings. As illustrations,
    molecular dynamics simulations are reported for the cases of a
    single ion imbedded in an electron plasma and for a two-component
    ion-electron plasma. Ion-ion, electron-ion, and electron-electron
    coupling effects are discussed for hydrogen-like Balmer alpha
    lines.
\end{abstract}

\pacs{52.65.Yy, 52.25.Vy, 05.10.-a}

\maketitle

\section{Introduction}
\label{secI}

    Theories for the interpretation of spectroscopic measurements in
    hot plasmas are crucial for diagnosing and studying laboratory and
    astrophysical plasmas. Well known theories such as the impact
    theory \cite{impact} or the unified theory \cite{unified} for
    electron broadening will be mentioned below. It should be
    emphasized that in this context a ``theory'' is a closed ensemble of
    many theoretical developments built on the basis of an initial set
    of hypotheses, namely a ``model" (possibly including a few
    variants). Within each theory, specific analytical and numerical
    methods are assumed for practical calculation of the physical
    properties that belong to that theory. The analysis here addresses
    the validity of common assumptions regarding the many-body Coulomb
    forces for computing radiative properties. More specifically, the
    lineshapes of ion emitters in fully ionized plasmas are
    investigated using classical molecular dynamics (MD) simulation.
    The advantage of MD simulation lies in its ability to account for
    all charge-charge interactions without approximation, providing a
    means to assess their relative importance for lineshape
    calculations.

    Most of the current models for the study of radiative properties
    of an emitter in a dilute medium treat the internal states of the
    emitter quantum mechanically, but use a semi-classical description
    for their interaction with the environment for the description of
    the environment. The system considered here is the same: a quantum
    emitter of charge $Ze$ imbedded in an equilibrium classical plasma
    with ions of charge $Z_{i}e$ and electrons of charge $-e$ at
    equilibrium. For simplicity, in all of the following it is assumed
    that the emitter couples to the plasma only through its monopole
    and dipole interactions.

    The line shape is then determined by  the
    Stark broadening of the total  electric field at the emitter.
    Three timescales have to be considered together: those
    characterizing ion motion, $\tau_{i}$, electron motion, $\tau_{e}$
    and a typical emitter dipole relaxation time, $\tau_d$. In all
    cases $\tau_{i}$ is much larger than $\tau_{e}$ due to their large
    mass ratio. Consequently, the ions do not respond directly to
    individual electron motions but rather to their averaged
    collective dynamics whose primary effect is to screen the ion-ion
    interactions. Conversely, the free electrons are coupled to all
    other electrons, and a slowly varying ion and emitter charge
    distribution through unscreened potentials. Similarly, the optical
    electrons bound to the emitter and responsible for radiation
    emission are coupled to the classical charges through a slowly
    varying screened ion field, and rapidly varying electric fields
    due to the colliding electrons via the dipole interaction. In some
    current theories the effects of this complex plasma environment
    are represented through the hypothesis of atomic or ionic emitters
    interacting with ions through a Coulomb field screened at the
    electronic Debye length and by free electrons moving on straight
    trajectories (no electron-emitter or electron-electron
    interactions) - hereafter called \textit{independent-electrons}.
    The simplest formulation of the impact theory for electron
    broadening is developed on this basis with three additional
    constraints for the independent electrons: 1) non-overlapping
    strong collisions (binary collision approximation), 2) positive
    lower bound for the impact parameter range (strong collision
    cut-off), and 3) a Debye screened Coulomb interaction between free
    and bound emitter electrons. In such theories, collective effects
    or correlations among the various charges are accounted for only
    in the static Debye screened ion-emitter and electron-emitter
    fields. The objective of our presentation here is to explore in a
    controlled fashion the effects of correlations, both static and
    dynamic, on properties relevant to plasma spectroscopy.

    Lineshape calculations follow two distinct stages: first, the
    ``stochastic" properties of the local electric field perturbation
    at the charged emitter from a classical plasma model are
    determined, and next these properties are used for modeling the
    dipole emission. The first point refers to a determination of
    charge structure and dynamics, and is essentially a nonlinear
    problem depending on the emitter monopole charge and the plasma
    parameters - a plasma physics problem. The second is connected to
    the behavior of a quantum emitter in a given time dependant
    stochastic perturbation - an atomic physics problem. The utility
    of MD simulation to treat the plasma physics component ``without
    approximation" is well-known \cite{plasmaMD}, although the
    application to study correlation effects of electrons with higher
    $Z$ ions is a new direction for hot plasmas \cite{tal1}. However,
    in order to assess the importance of such effects for
    spectroscopy, the atomic physics component is essential. The
    results presented here are a first attempt to combine MD
    simulation and atomic physics for a controlled description of both
    electron and ion broadening \cite{tal2}. Earlier applications of
    MD to spectral line broadening were limited to simulating the
    plasma ion fields \cite{ionMD}. Here, we emphasize the more
    difficult task of simulating as well the electrons which couple to
    the emitter in a qualitatively different way due to their
    attractive interaction.

    The classical plasmas considered for simulation are systems of
    point charges, either point electrons and the emitter monopole in
    a uniform positive background - hereafter called inhomogeneous
    jellium - or a two component plasma (TCP) of electrons plus the
    emitter monopoles. It is assumed that quantum mechanical effects are
    either negligible (as for the heavy ions and emitter) or can be
    incorporated through appropriate modifications of the classical
    Coulomb potential for the electrons. The latter is achieved by
    using a regularized Coulomb potential which remains finite at
    short distances, accounting simply for the true quantum
    diffraction effects \cite{Minoo}. This allows derivation of all
    the desired theoretical properties using the laws of classical
    mechanics, as well as the application of MD. A detailed study of
    the time independent statistical properties of an electron gas
    with an ion impurity has been carried out recently on the basis of
    such a regularized electron-ion potential \cite{tal1}. Two
    standard classical methods have been used for this study:
    molecular dynamics (MD) for numerical simulation for both
    dynamical and structural properties, and the hypernetted chain
    approximation (HNC) for the theoretical prediction of structural
    properties. This enables both cross comparisons and
    interpretations of results. An interpretation of the dynamical
    properties of the electron field autocorrelation function (FCF) at
    the impurity also has been carried out \cite{dufty-ilya} and is
    considered further here.

    Molecular dynamics is a unique way for providing realistic
    representations of the stochastic electric fields determining the
    dipole interaction of the emitter with its environment. Describing
    these fields by simulation allows one to obtain lineshapes
    accounting for all the correlations between charged particles.
    Such lineshapes would provide essential reference data to
    benchmark more efficient but phenomenological theoretical models
    developed for plasma diagnostics. Several recent theories for line
    broadening have been developed for investigations at new plasma
    conditions, including the hot and dense matter conditions needed
    for the presence of highly charged emitters. They suggest that
    part of the discrepancies found with experiments is an inadequate
    description of the emitter and electron perturber coupling
    mechanisms. The present study provides new information for this
    discussion.

    The types of questions posed are the following: 1)  What is the
    validity of straight line trajectories for the electrons in the
    presence of a charged emitter (neglect of electron-emitter,
    electron-electron, and electron-ion coupling in the dynamics)? 2)
    How well are the missing correlations of 1) captured by simple
    screening of the electron emitter interaction? 3)  What is the
    importance of multi-electron strong collisions? This study
    addresses these issues for two different classical plasma
    environments of the emitter. Both are well distinct from those
    resulting from independent electron models assumed for impact
    theories. The first and simplest is the inhomogeneous jellium used
    for describing electron broadening only. The corresponding study
    is hereafter referred to as the electron broadening theory (EBT).
    The second is a more realistic two component electron-ion plasma
    leading to a both ion and electron broadening theory (IEBT). The
    parameter range explored in this work is chosen to be compatible
    with hot and dense plasma diagnostics based on spectroscopy. These
    are conditions similar to those for which there are long standing
    discrepancies between theory and experiment. For a recent
    discussion and earlier references see reference \cite{3s3p}.

    In section \ref{secII}{} the main points of the classical plasma
    study for the positive charge emitter in the above two plasma
    environments is described. Section \ref{secIII}{} outlines the dipole
    relaxation stage of the
    broadening theories reported in this paper. Then, the main results
    obtained with the EBT, i.e. the effects of ion-electron
    correlations on electron broadening of spectral lines, are
    reported and discussed. Also in Section \ref{secIII}{} dipole
    relaxation in a
    TCP is considered and some preliminary results obtained using the
    the IEBT are shown. The examples reported below involve the Balmer
    alpha line for various hydrogen-like emitters of charge 1, 3, 5
    and 8 and various temperatures.

\section{Plasma environment modeling}
\label{secII}

    This section provides an illustration of the results that can be
    obtained from MD simulation for the two types of plasma environments
    considered for the emitter. In both the inhomogeneous jellium and
    the two component electron-ion plasma (TCP), the monopole of the
    emitter is included as a point charge. The dipole operator
    representing the internal states of the emitter is not considered in
    this section. For the density and temperature conditions considered
    the repulsive interactions of same sign charges effectively excludes
    pair configurations of the order of the DeBroglie wavelength and
    quantum effects are small. Consequently, the ion-ion, ion-emitter,
    and electron-electron interactions are taken to be Coulomb
    \begin{equation}
    \label{2.1}
        V_{12}(r)=Z_{1}Z_{2} e^{-r/\lambda}/r
    \end{equation}
    where $Z_{1}Z_{2}$ is positive. For practical purposes these Coulomb
    interactions have been screened at a distance $\lambda=s/2$ which is
    taken to be of the order of the system size $s$. This allows application
    of the usual periodic boundary conditions in the MD simulation.
    Since interactions among the charges introduce a physical screening
    at much shorter distances, this system size screening does not
    affect any of the properties being considered here. In contrast,
    electron-ion and electron-emitter interactions are attractive and
    the configurations within a DeBroglie wavelength become relevant. At
    such distances the Coulomb interaction must be modified within a
    classical description to account for quantum diffraction effects.
    There are many ways such quantum potentials can be constructed and
    here we use one of the simplest forms \cite{Minoo}
    \begin{equation}
    \label{2.1a}
        V_{ie}(r)=-Ze^{2}( 1-e^{-r/\delta }) e^{-r/\lambda}/r
    \end{equation}
    where $\delta =(2\pi \hbar ^{2}/m_{e}k_{B}T_{e})^{1/2}$ is the
    thermal De Broglie wavelength.
    This ion-electron potential regularization
    provides a well defined classical physics for opposite sign charge
    systems, and the various sophisticated classical N-body methods of
    classical statistical mechanics can be applied. The consequences
    and advantages of such a semi-classical approach for the radiative
    properties of ion emitters will be discussed in the last sections.

    Other parameters of interest are the average electron-electron
    distance $r_{0}=(3/4\pi N_{e})^{1/3}$ defined in terms of the
    electron density $N_{e}$, the electron thermal velocity $v_{0} =
    (2k_{B}T_{e}/m_{e})^{1/2}$, the mean electronic field
    $E_{0}=e/r_{0}^{2}$,
    the electron-electron coupling constant
    $\Gamma=e^{2}/r_{0}k_{B}T_{e}$, and the electron plasma frequency
    $\omega _{p}=(4\pi N_{e}e^{2}/m_{e})^{1/2}$. The coupling between the
    electrons and the impurity is measured by the ion-electron
    potential at the origin which, in units of $r_{0}$, becomes
    $\sigma =Z\Gamma /\delta $. Similar quantities are defined for the
    ions in the TCP by replacing the magnitude of the electron charge
    and density by the corresponding ion values.

    Molecular dynamics simulations are carried out using a number $N$ of
    electrons for the jellium, or $N$ electrons and $N/Z_{i}$ ions for
    the TCP. In the first case a point charge ${Z\ll N}$ representing the
    emitter is included while for the TCP the ions are considered as the
    emitters for simplicity. The simulation is done using periodic
    boundary conditions on an elementary cubic cell of size $c$ large
    enough to ensure $\lambda>\lambda _{D}$, where $\lambda _{D}$ is the
    Debye wavelength, and $\lambda \leq s/2$. These conditions assure
    that the physical screening due to the charge interactions occur at
    a shorter scale than the system size screening, as noted above.
    Because $N$ must be large to agree with the condition $Z\ll N$ the
    impurity-impurity correlation inherent to periodic boundary
    conditions is negligible. Moreover as we are only interested in the
    charge structure and dynamics in the ion vicinity the usual Ewald's
    sum is not required. Particle motion in MD simulation is achieved
    using a Verlet algorithm with a time step error of order $(\Delta
    t)^{4}$:
    $\mathbf{r}_{n+1}=2\mathbf{r}_{n}-\mathbf{r}_{n-1}+\mathbf{a}_{n}(\Delta
    t)^{2}$.

\subsection{Ion emitter in an electron gas}
\label{subsecIIa}

    This section summarizes the study of a single ion impurity (the
    monopole of the emitter) imbedded in an electron gas. A uniform
    positive background maintains overall charge neutrality but has no
    effect on particle motion since, according to periodic boundary
    conditions used in molecular dynamics, the charges move across an
    infinite system. The motivations for doing the impurity case are:
    1) it is a well defined and simpler problem than the TCP case in
    comparison, and 2) all the existing theories for electron
    broadening to compare with are based on an impurity approach. The
    main statistical properties of the electron structure close to the
    ion (e.g., electron density, electric field probability
    distribution) are derived using both MD simulation and the
    classical hypernetted chain (HNC) integral equations for the
    impurity-electron correlation function.

    Ion-electron and electron-electron coupling affect both the static
    and dynamic properties of the electrons near the impurity ion. The
    former determines the dominant $Z$ dependence of these properties,
    while the latter controls the screening and other quantitative
    effects. The electron density, electric microfield distribution,
    and electric field autocorrelation function have been studied for
    various temperatures, densities, and impurity charge number $Z$.

\begin{figure}[ht]
    \includegraphics[width=10cm]{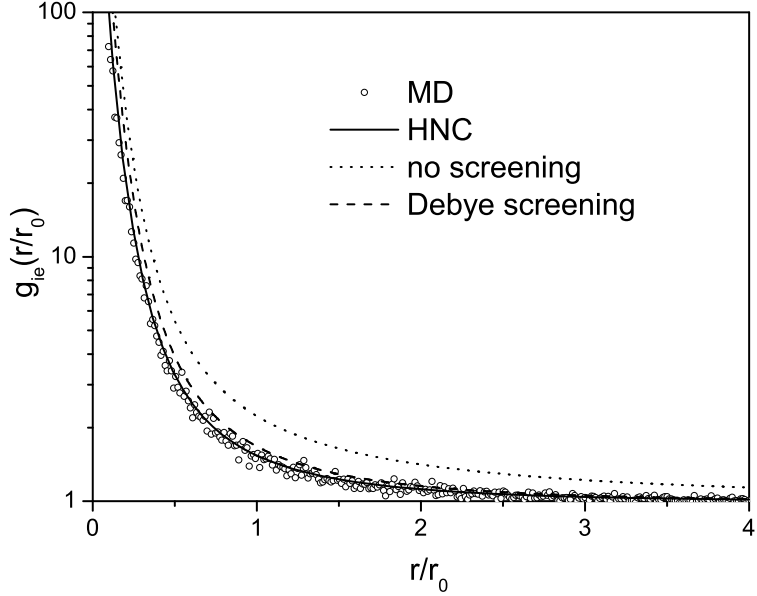}
    \caption{%
        Ion-electron pair correlation function at $N_{e}=10^{19}
    cm^{-3}$, $T_{e}=50000K$, and {Z=8}.
    Solid line: HNC;
    circles: MD;
    dashed: no e-e coupling and Debye screening;
    dotted: no e-e coupling and no screening.}
    \label{fig1}
\end{figure}

\begin{figure}[ht]
    \includegraphics[width=10cm]{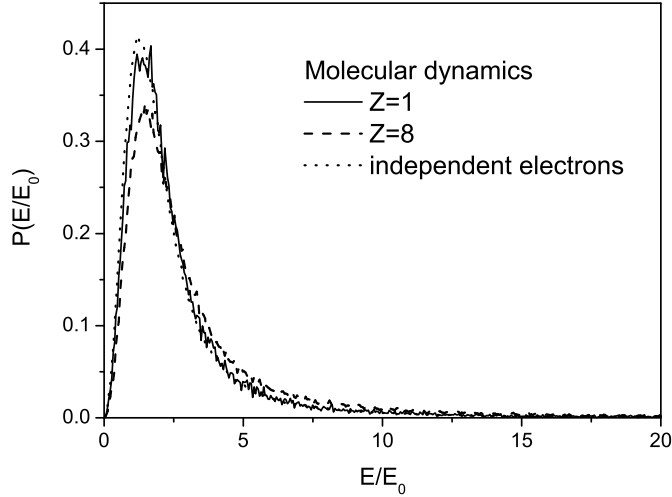}
    \caption{%
        Field distribution at $N_{e}=10^{19} cm^{-3}$, $T_{e}=200000 K$.
    Solid: $Z=1$;
    dashed: $Z=8$;
    dots: independent electrons.}
    \label{fig2}
\end{figure}

    Consider first the electron density near the impurity. The
    simplest prediction is given by the Debye-Huckel approximation
    leading to an enhanced density near the ion and Debye screening at
    large distances. This approximation is expected to fail for
    increased electron-ion coupling (larger values of $Z$) and/or
    increased electron-electron coupling. In that case an adequate
    theory is provided by the two component HNC equations
    \cite{Rogers} specialized for the single impurity case, i.e. for a
    vanishing ion density. The impurity-electron pair distribution
    function $g_{ie}(r)$ is obtained by numerical solution to these
    equations as a function of the direct electron-electron
    correlation function $c_{ee}(r)$ calculated separately. Aside from
    normalization, this pair correlation function is the average
    electron number density, $n_{e}\left( r\right) =n_{e}g_{ie}(r)$ at
    a distance $r$ from the ion. Both MD simulation and the HNC
    approximation are restricted to a limited range of values for the
    impurity charge, density, and temperature. Outside of this range
    frequent electron trapping and non-convergent iterative solution
    methods for the HNC equations occur invalidating the results.
    Within the parameter space considered here MD and HNC results are
    in good agreement as illustrated in Fig. \ref{fig1}.
    The enhanced electron accretion around the ion impurity is due to
    the attractive impurity-electron interaction. If this coupling is
    neglected the electron distribution becomes uniform regardless of
    the electron-electron coupling. Clearly, the independent electron
    model is a bad approximation for the electron density except for
    $Z=0$. Also shown in Fig. \ref{fig1}{} is the result obtained by
    including the impurity-electron interaction but neglecting
    electron-electron interactions. This misses the screening effects
    at larger distances due to the electron coupling. For the
    conditions considered, the electron coupling is small,
    approximately $\Gamma = 0.116$,
    and the simple addition of Debye screening
    gives quite good results. Other properties involving the local
    electron electric field at the impurity are of primary interest
    for spectroscopy since they determine the dipole coupling to the
    internal states of the emitter. The electric field for the current
    case is determined from the gradient of the potential in Eq.
    \ref{2.1a}. For the independent electron approximation, the
    screening of this field is changed to the Debye length.
    Theoretical predictions of the electric microfield distribution
    can be obtained in terms of $g_{ie}(r)$ and the results using HNC
    have been compared with those from MD, again with good agreement.
    The MD results are illustrated in Fig. \ref{fig2}{} for two values
    of impurity charge $Z=1$ and $Z=8$.
    The effect of increased charge number is to decrease and broaden
    the distribution. Also shown is the result for independent
    electrons. This is the $Z=0$ interaction without electron-electron
    interactions, but with Debye screened fields. It is seen to be
    similar to the $Z=1$ results. As with the average electron
    density, the dominant Z dependence is due to the impurity-electron
    coupling. However, at fixed $Z$ the electron-electron coupling is
    responsible for important quantitative changes in the location of
    the maximum and width of the distribution.

\begin{figure}[ht]
    \includegraphics[width=10cm]{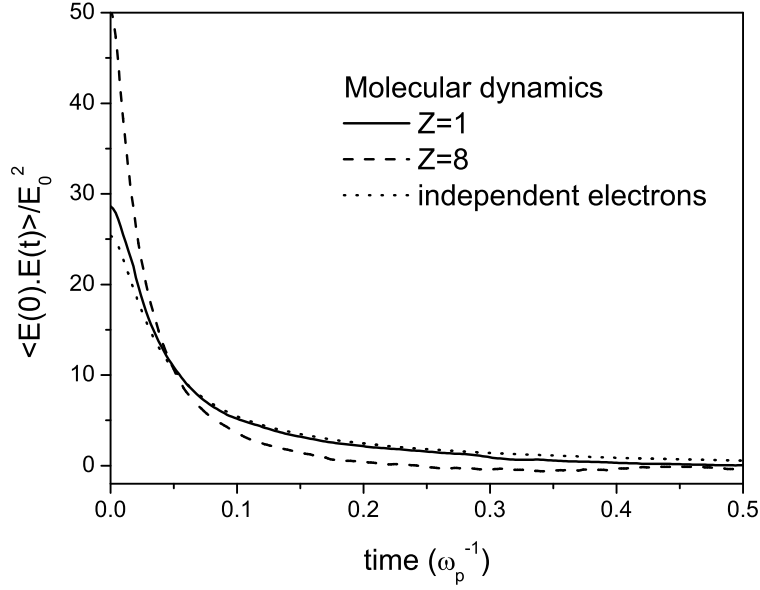}
    \caption{%
        Field autocorrelation function, same as Fig. \ref{fig2}.}
    \label{fig3}
\end{figure}

    The simplest measure of the electric field dynamics is  given  by
    the field autocorrelation function (FCF). Figure \ref{fig3}{} shows
    the time dependence for the cases $Z=1$ and $Z=8$ at
    $T_{e}=200000K$.
    Also shown is the independent-electron limit obtained for screened
    fields but removing all interactions. As with the field
    distribution function, the result is similar to the {Z=1} case.
    The ad hoc screening of the fields assures that the initial value
    is similar to that with electron-electron coupling. This screening
    also assures that the straight trajectory dynamics is effectively
    restricted to a Debye sphere. However, the electron-impurity
    coupling changes this agreement dramatically at larger values of
    $Z$. Qualitatively, the behavior of the correlation function can
    be summarized as follows: For increasing impurity charge number
    the initial value $<E^{2}>/E_{0}^{2}$ increases while the
    correlation time decreases. Both effects are missed by the
    independent electron approximation, but are regained qualitatively
    if only impurity-electron coupling is retained (without
    electron-electron interactions). However electron-electron
    coupling is required for quantitative effects as indicated below.

\begin{figure}[ht]
    \includegraphics[width=10cm]{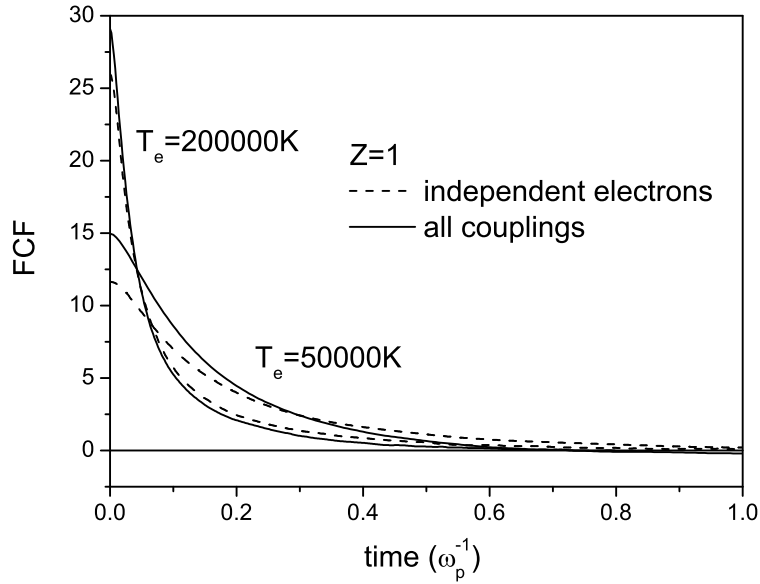}
    \caption{%
        FCF for two temperatures at Z=1.
    Dash: no coupling;
    solid: all couplings.}
    \label{fig4}
\end{figure}

    Now consider the effects of electron-electron coupling in more
    detail for the case $Z=1$. Figure \ref{fig4}{} shows the results for
    two different temperatures, $T_{e}=200000K$ and $T_{e}=50000K$.
    This corresponds to two different values for the electron-electron
    coupling constant, $\Gamma =0.029$ and $\Gamma =0.116$, and two
    different values for the electron-ion coupling constant,
    $\sigma =0.5$ and $\sigma =1$, for the high and low temperature
    cases, respectively. Also shown for comparison are the results for
    the independent electron model. The initial value of the FCF is seen
    to increase at constant $Z$ for increasing temperature. This can
    be understood since the regularized potential leads to a
    covariance proportional to $1/\delta \sim{T^{1/2}}$. The
    correlation time decreases at higher temperatures, in part at
    least due to the increased thermal velocity. As expected, the
    deviation of the independent electron model from these results
    increases with increased electron coupling. The apparent agreement
    for the high temperature case is misleading in this presentation,
    since the percent discrepancies are large over the entire time
    interval as will be shown in Fig. \ref{fig7} below. The effect of
    electron-electron coupling on the dynamics is therefore important.
    This is more evident for the lower temperature comparison.

\begin{figure}[ht]
    \includegraphics[width=10cm]{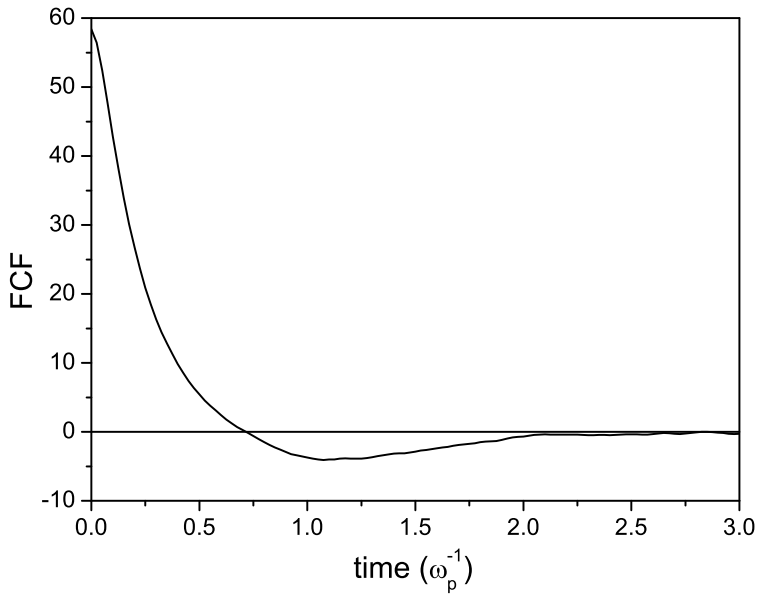}
    \caption{FCF anti-correlation for $Z=8$ and $T_{e}=100000K$.}
    \label{fig5}
\end{figure}

\begin{figure}[ht]
    \includegraphics[width=10cm]{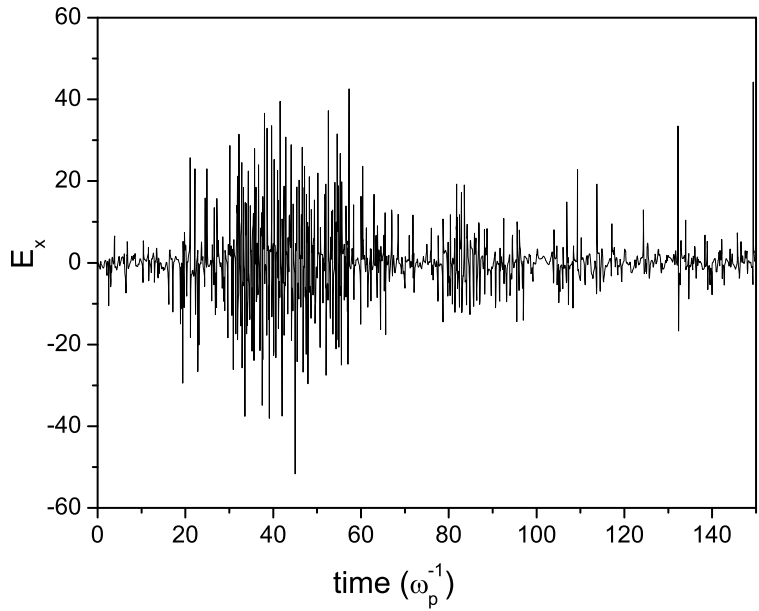}
    \caption{%
        Example of temporarily oscillating electric field at the
    ion. $Z=8$ and $T_{e}=100000K$.}
    \label{fig6}
\end{figure}

    The strong electron-impurity coupling case of $Z=8$ is shown in
    more detail in Fig. \ref{fig5}.
    A striking new feature is the anticorrelation (negative FCF) at
    intermediate times. This effect increases for larger $Z$ and also
    for decreasing temperature, corresponding to the enhanced
    electron--ion interactions. This increases the probability of electron
    configurations near the ion. Many of these lead to temporary
    orbiting trajectories with associated large oscillatory electric
    fields. When such trajectories dominate the contribution to the
    FCF, their changing fields on average will lead to an
    anticorrelation as observed in Fig. \ref{fig5}. The occurrence of
    such events is easily seen in the MD simulations, as shown in Fig.
    \ref{fig6}. The total field is dominated by the oscillatory fields
    of a few close orbits. These oscillations persist on the time
    scale of the FCF, but are seen to break up on a time scale an
    order of magnitude longer, due to electron-electron interactions.
    Thus, the effect of anticorrelation is due to quasi-bound orbits,
    but their correct equilibrium statistical sampling requires the
    electron-electron interactions to assure the properly weighted
    ensemble of $\it{temporary}$ orbiting trajectories.

    A general remark can be addressed about the MD approach in the
    context of this discussion. The field history sampling is started
    after a thermal equilibration stage intended to reach a stationary
    energy state. Independent samplings are obtained starting from a
    new uniform random configuration of particles followed by a
    thermal equilibration stage. This process is repeated many times
    in order to build a very large sample set. If the
    electron-electron coupling is neglected, the equilibration stage
    does not favor the electron accretion around the ion impurity in
    the correct way. As a result the sample set obtained is certainly
    not the same as the sample set obtained when the electron-electron
    interaction is accounted for. An improved sampling could be built
    using initial configurations chosen in order to agree with the
    nonlinear electron distribution around a positive charge. This way
    has been followed to develop a simple theoretical model for the
    FCF \cite{dufty-ilya}.

    It will be seen later that the spectral line width is proportional
    to the time integral of the FCF. It is difficult to predict the
    $Z$ dependence of this integral a priori due to the above
    competing effects of increasing initial value and decreasing
    correlation time. Consider again the FCF for $Z=1$ in Fig.
    \ref{fig4}{} at two different temperatures. The time integral of
    these correlation functions and another at $100000K$ are shown as
    a function of temperature in Fig. \ref{fig7}. Also shown are the
    corresponding results for the independent particle model.
\begin{figure}[ht]
    \includegraphics[width=10cm]{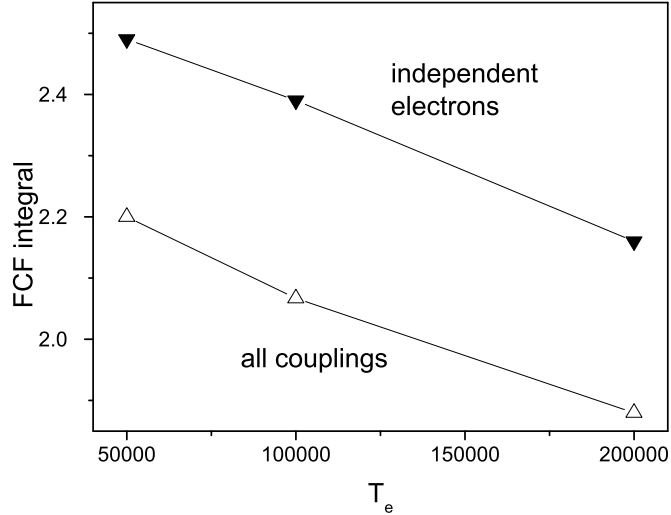}
    \caption{%
        FCF integral versus temperature for $Z=1$. white
    triangles: all couplings, black triangles: independent-electrons.}
    \label{fig7}
\end{figure}
    It is seen that the FCF integral decreases with temperature, so
    evidently the dynamical effect of a decreasing correlation time
    dominates the increasing initial correlations. The effects of
    electron-impurity and electron-electron coupling leads to an
    approximately $15$ percent decrease in the integral at all
    temperatures relative to that for the independent electron model.

\begin{figure}[ht]
    \includegraphics[width=10cm]{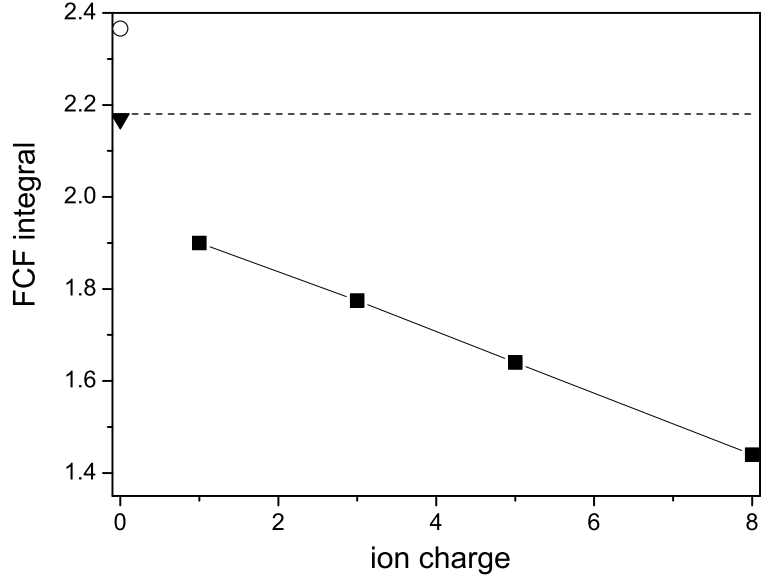}
    \caption{%
        FCF integral versus emitter charge at $T_{e}=200000K$.
    Squares: all couplings;
    dashed: independent-electrons;
    triangle: all couplings with $Z=0$;
    circle: unscreened uncoupled electrons with $Z=0$.}
    \label{fig8}
\end{figure}

    Finally, the $Z$ dependence of the FCF integral is demonstrated in
    Fig. \ref{fig8}. Again dynamical effects of a decreasing
    correlation time and anticorrelation dominate, leading to an
    approximately linear decrease of the integral with increasing $Z$.
    Consider first the value $Z=0$ (no impurity-electron coupling).
    The triangle represents the results with electron-electron
    coupling, while the line is the independent electron model. It is
    seen that the use of the screened field in the latter accounts
    well for the relevant electron-electron coupling. This is
    consistent with the discussion following Fig. \ref{fig3}. Also
    shown is the independent electron model without using the screened
    fields (open circle). This shows clearly that electron-electron
    coupling is almost a {10} percent effect at $Z=0$. This is
    significant since the electron-electron coupling constant is quite
    small ($\Gamma =0.029$). For $Z=1$ the electron-emitter
    interaction is quite important, decreasing the integral by almost
    $15$ percent below the independent electron value. This coupling
    continues to dominate for increasing $Z$. As noted above, for
    $Z>0$ there is an additional importance of the electron-electron
    coupling in the MD which is needed to establish the correct weight
    for the increasing number of quasi-bound orbits.

\subsection{Ion-electron two component plasma MD simulations}
\label{subsecIIb}

    In this subsection MD studies of an equilibrium, charge neutral
    two-component plasmas (TCP) of $N$ electrons and $N/Z_{i}$ ions
    are reported. Charge neutrality is guaranteed by the charge
    concentration. Preliminary classical TCP MD simulations in this
    context of applications to spectroscopy have been reported
    elsewhere \cite{tal2}. The challenging aspect of such simulations
    is to move slow and fast particles at the same time. Such a study
    of fast and slow processes requires simulations stable over a long
    period of time, long enough to account for ion motion and based on
    a time step compatible with electron motion. This problem has been
    addressed mostly in the context of hydrogenic plasmas
    \cite{hydrogen} and is extended here to higher $Z$ ions. The
    emitter is taken to be one of the ions of the TCP.

    It is worth noting at this point the distinction between classical
    MD applied to the TCP, where both ions and electrons move
    dynamically, and ``quantum MD" where the ions move classically on a
    self-consistent potential energy surface of the electrons
    \cite{Collins,dharma}. In the latter, at each time step for the ions
    the ground state configuration for the electrons is determined by
    some quantum method (e.g., density functional theory). Subsequently
    the forces between the ions in this electronic configuration are
    calculated and the ions are moved along the corresponding classical
    trajectory. The process is repeated at the next time step. The two
    classes of simulations appear complementary, with quantum MD being
    more appropriate for lower temperatures and semi-classical MD more
    appropriate at higher temperatures.

\begin{figure}[ht]
    \includegraphics[width=10cm]{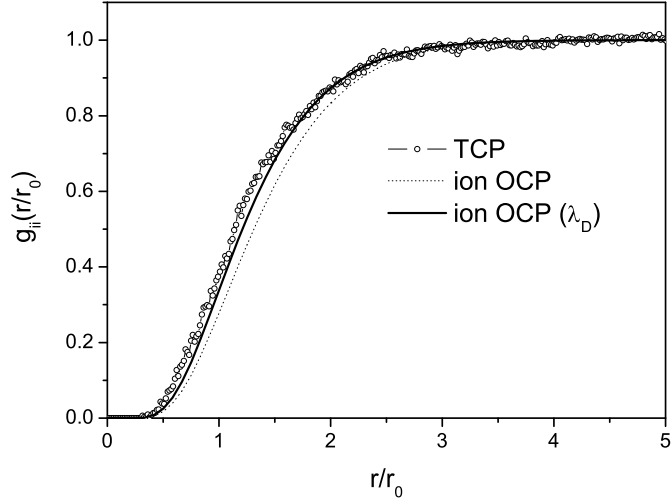}
    \caption{%
        Pair correlation function.
    Solid: OCP;
    circles: TCP;
    dots: no screening.}
    \label{fig9}
\end{figure}

    We first verify that the MD simulation properly describes the
    expected ion-electron screening mechanism. For this purpose the
    simulation of a neutral mixture of hydrogen-like carbon ions $Z=5$
    in electrons has been carried out with the temperature
    $T_{e}=50000K$ and the electron density $N_{e}=10^{19}cm^{-3}$. As
    discussed following Eq. \ref{2.1}, a large cutoff length $\lambda$
    comparable to the system size has been used in the potential to allow
    the periodic boundary conditions. Since the physical screening
    resulting from the collective effects of these ``Coulomb"
    interactions is at a much smaller scale, this cutoff does not
    affect the study of that screening. A comparison of the ion-ion
    pair distribution function for the TCP with ``Coulomb" interactions
    with that for a corresponding ion OCP using Debye screened
    interaction is made. In the OCP system there are no electrons and
    ion-ion Coulomb potential is screened at the Debye length
    phenomenologically, while such screening must be generated
    directly by the electrons in the TCP simulations. Figure
    \ref{fig9}{} shows the rather good agreement between the two
    simulations, giving an indirect evidence of the effectiveness of
    the ion-ion dynamic screening by the electrons (also shown is the
    OCP without screening to quantify the effect being studied). The
    TCP result shows a small enhancement of the screening mechanism
    over that of the OCP at shorter distances. This behavior suggests
    that the short range ion-ion interaction could be well described
    using an effective ion charge $\overline{Z}<Z$. An rough estimate
    not shown gives $\overline{Z}\simeq 4.7$.

\begin{figure}[ht]
    \includegraphics[width=10cm]{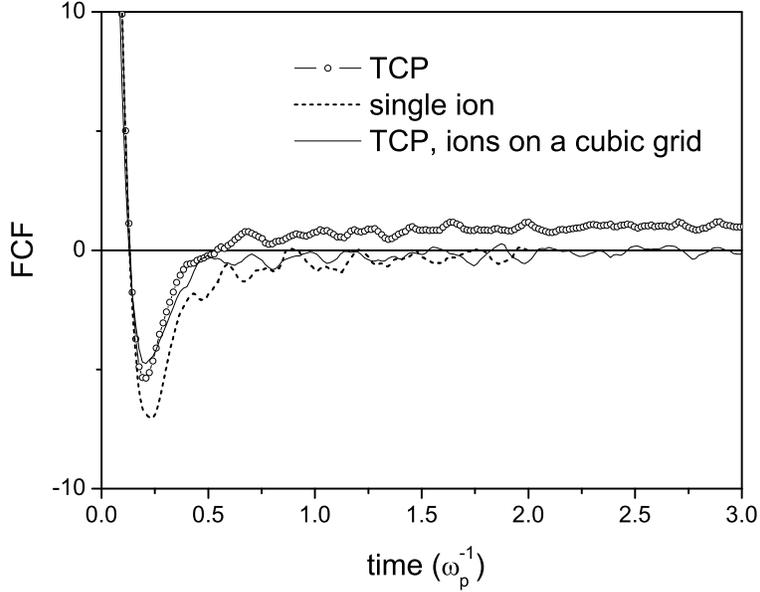}
    \caption{%
        Electron field autocorrelation function.
    Dashed: impurity;
    circles: TCP;
    solid: cubic array.}
    \label{fig10}
\end{figure}

\begin{figure}[ht]
    \includegraphics[width=10cm]{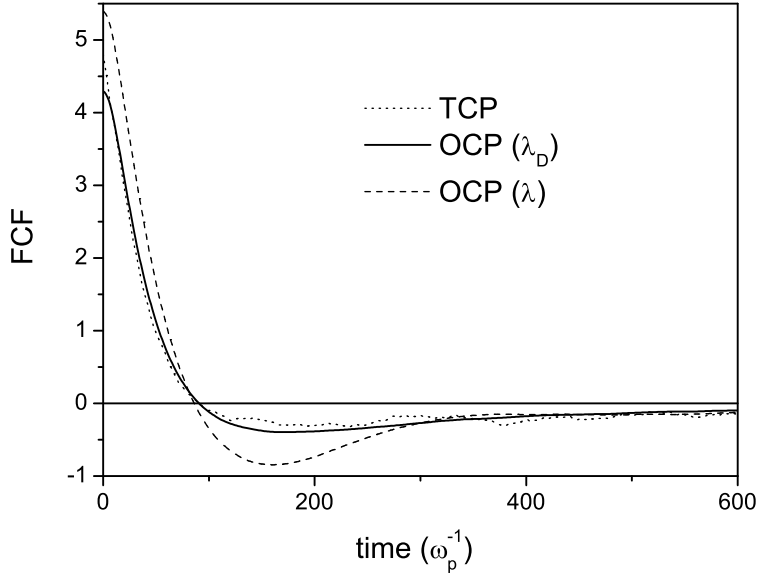}
    \caption{%
        Total field autocorrelation function.
    Dots: TCP simulation;
    solid: ion OCP simulation;
    dashed: ion OCP no screening.}
    \label{fig11}
\end{figure}

    Evidence of screening also can be demonstrated via the dynamics of
    the total field due to both ions and electrons at one of the ions
    (considered to be the emitter). MD simulation allows determination
    of statistical data for each particle type involved. In the
    electron-ion TCP two kinds of time dependencies coexist, the high
    frequency and the low frequency dynamics related to electron
    motion and ion motion, respectively. Both of them are of interest
    for probing or discussing the simulation procedure accuracy. For
    the high frequency data the electron field autocorrelation
    function at the ion has been calculated. This is similar to the
    property calculated in the subsection above for jellium, except
    that now it is in the presence of many other ions. Figure
    \ref{fig10}{} shows a comparison of the field autocorrelation
    function for
    the field seen by
    a single ion in electrons (jellium) and for the TCP.
    Despite some noise due to insufficient sampling it can be observed
    that for the TCP the field correlation function does not go to $0$
    as for the impurity in jellium case. The timescale used in this
    simulation is the inverse electron plasma frequency, compatible with
    the high frequency mechanisms of the electrons. On this time scale
    the electron FCF tends to a constant value rather than zero,
    denoting the occurrence of a low-frequency component in the electron
    field, due to its coupling to the slower ion dynamics. Contrary to
    the impurity ion in jellium, a given ion is no longer a center of
    symmetry for the average electron structure. Rather, this average
    electron structure is a blurred image of the random ion
    configuration which remains quasi-static at this time scale. This
    low frequency component disappears, as shown on Fig. \ref{fig11}, if
    the ions are located on the nodes of a cubic array where each ion is
    a center of symmetry of the infinite system. As expected, on the
    time scale of this low-frequency dynamics this seemingly constant
    value goes to $0$.

    In many theoretical formulations of lineshapes, such as the fast
    fluctuation approximation of the next section, it is assumed that
    the electron electric field relaxes quickly relative to the time
    scale of the emission or absorption process. One consequence of
    the occurrence of a low-frequency component in the TCP electron
    field is that these approximations are no longer valid.

    The FCFs generated by a one component plasma of ions interacting
    through a Coulomb potential 1) with a screening length $\lambda
    _{D}=$ Debye length, 2) with a box screening length
    $\lambda>\lambda _{D}$ are compared with the FCF of the low
    frequency field component of the ion plus electron total field on
    Fig. \ref{fig11}.
    The good agreement of the two curves and the large discrepancy with
    the no screening result can be interpreted again as the proof that
    the screening mechanisms take place properly in the TCP molecular
    dynamics simulations.

\section{Electron-ion coupling effects on lineshape}
\label{secIII}

\subsection{Dipole relaxation mechanisms}
\label{subsecIIIa}

    Linear response methods can be
    applied to the dipole relaxation of a quantum emitter in a
    classical plasma environment \cite{griem}. According to the energy
    involved it is postulated that the perturbation by the environment
    never induces transitions between the two manifolds for which the
    transition is considered within the characteristic time of the
    emission - no quenching approximation. This approximation
    therefore neglects emitter-bath coupling via the radiated dipole
    transition and any contribution of the dipole transition moment at
    equilibrium. As a result, the spectral profiles of dipole
    spontaneous-emission are given in terms of the autocorrelation
    function $C(t)$ of the system dipole moment $\mathbf{q}(t)$, via a
    Fourier transformation:
    \begin{eqnarray}
        I(\omega) &=& {\rm Re} \frac{1}{\pi} \int_{0}^{\infty}
                     dt~ C(t)\, e^{{\rm i} \omega t}\,,\\
        C(t) &=& {\rm tr}\;\{U^{*}(t)\rho\,\mathbf{q}U(t)\cdot\mathbf{q}\}
    \end{eqnarray}
    The trace is taken over an equilibrium Gibbs statistical ensemble
    $\rho$ for the total system of plasma plus emitter. The
    Hamiltonian is taken to be the sum of the plasma Hamiltonian
    (including the point monopole of the emitter) $H_{p}$, the
    Hamiltonian for the internal states of the emitter $H_{e}$, and
    their interaction due to the plasma electric field coupling to the
    internal emitter dipole. For simplicity here, the Doppler
    broadening due to emitter center of mass motion has been assumed
    to be statistically independent of this dipole broadening. The
    operator $\mathbf{q}(t)=U(t)\mathbf{q}U^{*}(t)$ obeys the
    Heisemberg equation
    \begin{equation}\label{3.1}
        {\rm i}\hbar\frac{d\mathbf{q}(t)}{dt} =
        [H_{e}-\mathbf{E}(t)\cdot\mathbf{d},\mathbf{q }(t)]
    \end{equation}
    The second term in the commutator describes the perturbation due
    to the total electric field of the plasma $\mathbf{E}(t)$. Due to
    the no quenching approximation, $\mathbf{d}$ is restricted here to
    the part of the internal emitter dipole operator coupled to this
    field. The time dependence of $\mathbf{E}(t)$ is generated by the
    plasma Hamiltonian alone. This shows most directly the utility of
    MD simulation in addressing the most difficult part of a line
    broadening problem: identifying properly and completely the
    environment of the emitter. In the present case this is given by
    an electric field history. To represent an equilibrium ensemble
    for the environment a collection of field histories for different
    initial conditions is given, the Schr\"odinger equation solved in
    each case, and a simple algebraic average is performed over all
    such solutions to get the line profile. In practice the numerical
    solution to the Schr\"odinger equation can be straightforward or
    complicated depending on the electric field history. This is the
    case for high $Z$ radiators for which the quasi bound orbits
    described above can give rise to high frequency, strong fields
    (e.g., as in Fig. \ref{fig6}). The calculations require a fast
    integration process based on a polynomial development using the so
    called Euler-Rodrigues coefficients. A detailed presentation of
    the method applied to hydrogenic lines has been given elsewhere
    \cite{marco}.

    The existing theoretical analysis generally entails further
    approximations not necessary in the above MD/Schr\"odinger equation
    simulation. Due to the small electron-ion mass ratio, Stark
    effects resulting from slow ion and fast electron micro-fields are
    generally considered separately. The Stark effect due to ions is
    approximated by a static electric field, while electron broadening
    is described as a dynamical process resulting from the high
    frequency fluctuation of the electron electric field. The usual
    approach for electron broadening \cite{griem} for ion emitters in
    plasmas is a binary collision model involving independent
    electrons moving on straight trajectories at constant velocity.
    Electron correlations are accounted for only indirectly by
    screening the electron-emitter potential at the Debye length.
    There are other theoretical models with different approximations,
    but for the discussion here we will primarily compare simulations
    for two cases: 1) the electrons are considered as free particles
    that do not interact with each other but perturb the ion emitters
    with a field obtained from the potential written in Eq.
    (\ref{2.1a}) with $\lambda =\lambda _{D}$; and 2) the electrons
    have Coulomb interactions with each other and couple to the
    emitter with a field obtained using Eq. (\ref{2.1a}) with $\lambda
    =s/2$. The latter, of course, is the correct treatment of
    correlations.

    A more compact formulation of the problem is obtained using a
    Liouville operator representation \cite{fano} allowing to
    disentangle the evolution operator from the internal emitter dipole
    operator $\mathbf{q}$. Liouville operators are defined according to
    the following relation:
    $ L_{0}\equiv \frac{1}{{\rm i}\hbar}[H_{0},\cdot]$.
    In this representation the dipole correlation function can be
    rewritten $C(t)=Tr(\bf{q}\cdot U(t)\rho\bf{q})$. Equation \ref{3.1}
    is replaced by the following stochastic equation for the evolution
    operator alone:
    \begin{eqnarray}
        \frac{dU_{\alpha}(t)}{dt} &=& (L_{0}+\mathbf{D}\cdot
        \mathbf{E}_{\alpha}(t))U_{\alpha}(t),\quad U_{\alpha}(0)=1
        \nonumber\\
        \textbf{U}(t) &=& \{U_{\alpha}(t)\}_{av}
        =\frac{1}{N}\sum_{\alpha=1}^{N}U_{f_{\alpha}}(t)
    \end{eqnarray}
    where $\mathbf{E}_{\alpha}(t)$ is one of the electric field
    histories, $\mathbf{D}$ is the Liouville representation of
    $\mathbf{d}$ and $\mathbf{U}(t)$ denotes the solution of the
    stochastic equation.

    A useful approximation described in Appendix A, the fast
    fluctuation limit, can be obtained when the characteristic time
    for the field correlation function is much smaller than the
    typical relaxation time of the physical process investigated -
    here the dipole \cite{lineshape}. In this limit the stochastic
    equation reduces to:
    \begin{equation}
        \frac{d\textbf{U}(t)}{dt}=(L_{0}-\textbf{W})\textbf{U}(t)
    \end{equation}
    where $\textbf{W}$ is proportional to the field correlation
    function integral
    $\int_{0}^{\infty}\{\mathbf{E}_{\alpha}(0)\cdot
    \mathbf{E}_{\alpha}(t)\}_{\alpha}dt$.
    Thus questions regarding the
    effects of charge-charge coupling on the linewidth can be
    addressed directly from the simulation of the field
    autocorrelation function as described in the previous sections.

\subsection{Electron broadening}
\label{subsecIIIb}

    The purpose of this section is to apply the EBT developed herein
    for discussing the occurrence of charge-charge coupling effects in
    electron broadening. This issue is motivated in part by recent
    experimental and theoretical studies performed on a series of
    lines emitted by ions of increasing charge \cite{3s3p}. The
    specificity of these lines relies on the absence of ion
    broadening. However, in contrast, the present study concerns lines
    with ion broadening so the generalization of the conclusions here
    to all kinds of lines would be inappropriate. Ion lines with non
    negligible ion broadening have been selected as they fit also the
    objective of studying Stark broadening in  two component plasmas
    accounting for complete charge-charge coupling.

    According to the EBT, the study of electron broadening
    relies first on the construction of large field
    sample-sets at the emitter generated by molecular dynamics. Then,
    the lineshape or simply the full width at half maximum (FWHM),
    $\gamma(Z,T_{e})$, is obtained after numerical solutions of
    the Schr\"odinger equation and
    its average. Due to the various characteristic times of the
    problem, a given field history can be split into a number of
    independent samples depending on the final purpose. In the present
    case, the dipole relaxation simulation requires samples much more
    longer than those necessary to obtain the field statistical
    properties. In addition, the study of electron broadening as a
    function of the emitter charge can be limited by calculation
    costs, since the dipole response of the emitter decreases as the
    inverse of the square of the emitter charge. For instance, to
    carry out an equivalent study with respect to the noise inherent
    to this kind of statistical average for emitters of net charge
    $Z=1$ and $Z=8$, it would be necessary to generate a field sample
    set and then integrate the Schr\"odinger equation over a time about
    $20$ times longer for $Z=8$ than for case $Z=1$. Accordingly, the
    present study is limited to cases relevant for the fast
    fluctuation limit (FFL) allowing a simpler theoretical
    representation of the line width and thereby avoiding the full
    integration process. It should be noted that, for a given
    temperature, if conditions for the FFL are fulfilled for a chosen
    line in the case $Z=1$ they are even more relevant for the same
    line in the case $Z=8$. However, the dipole autocorrelation will
    be calculated by complete integration of the Schr\"odinger equation
    in a few cases in order to check the FFL and for cross
    comparisons. In order to probe the statistical properties of the
    local microfield, Balmer-alpha lines for hydrogen like emitters of
    charge $Z=1,3,5,8$ have been chosen together with the three plasma
    temperatures $T=5 \times 10^{4},10^{5},2\times 10^{5}K$ selected
    for this study. Both the dipole moments of the lower and the upper
    level manifolds are coupled to the perturbing field giving rise to
    the relaxation process and line broadening of the emitted dipole
    transition $n=3 \rightarrow n=2$. Qualitatively, for the Balmer
    alpha line, the electric field perturbation induces two opposite
    broadening mechanisms, i.e. the line broadening increases with the
    covariance characteristic of the field strength while, due to some
    time averaging inherent to the dipole relaxation process, it
    decreases with an increasing fluctuation dynamics of the
    perturbation.

\begin{table}
    \caption{%
        \label{tab1}
    Balmer alpha FWHM $(s^{-1})$ versus temperature for $Z=1$.}
    \begin{ruledtabular}
    \begin{tabular}{lcr}
    Temperature&$\gamma_{FFL}$&$\gamma_{SE}$\\
    \hline
    $5 \times 10^{4}K$  & $10.7 \times 10^{12}$ & $11.8 \times 10^{12}$\\
    $10^{5}K$  & $9.88 \times 10^{12}$ & $10.2 \times10^{12}$\\
    $2\times 10^{5}K$ & $9.32 \times 10^{12}$ & $9.93 \times 10^{12}$\\
    \end{tabular}
    \end{ruledtabular}
\end{table}

    Comparisons of the FWHM are reported in Tab.~\ref{tab1}, where
    $\gamma_{FFL}$ is the fast fluctuation limit result and
    $\gamma_{SE}$ results from simulation i.e. from a full integration
    of the Schr\"odinger equation. The discrepancy is smaller than $10
    \%$ of the FWHM. In addition the exponential decay of $C(t)$
    predicted by the FFL also is confirmed for the simulations (not
    shown).

\begin{figure}[t]
    \includegraphics[width=10cm]{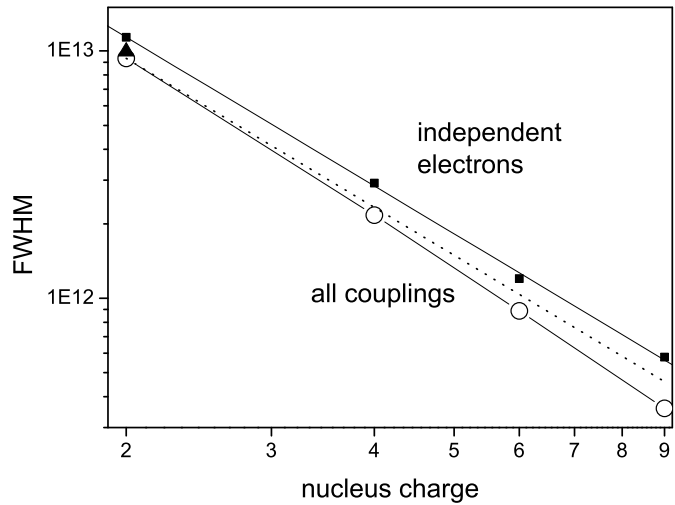}
    \caption{%
        Balmer-alpha FWHM, $Z=1$, $T_{e}=200000$ K.
    Squares: independent-electrons;
    solid line: independent-electron linear fit;
    circles: charge-charge coupling;
    black triangle: complete integration;
    dots: $1/(Z+1)^{2}$ line.}
    \label{fig12}
\end{figure}

    For the high temperature case, these results are plotted on a log
    scale in Fig. \ref{fig12}, versus the nucleus charge, together
    with the corresponding independent electron calculations (free
    electrons moving on straight trajectories that perturb the emitter
    via the potential defined in Eq.~\ref{2.1a}{} screened at the Debye
    length). The field sampling generation cost by the MD technique is
    quite low for noninteracting particles. Therefore, these
    calculations can be performed by integration of the stochastic
    equation with a satisfactory noise level. The FWHM calculations
    fit the $1/(Z+1)^{2}$ behavior even without coupling to the
    emitter due to the atomic physics dependence (dipole matrix
    elements) of the linewidth. These calculations confirm the
    capability of the simulation technique to provide relevant results
    even in the fast fluctuation limit domain. The entire discrepancy
    between the two results of Fig. \ref{fig12}{} is therefore a
    consequence of the charge-charge coupling in the time integral of
    the FCF as a function of the emitter charge (i.e., the effect
    discussed in Fig. \ref{fig8}). This discrepancy increases with
    decreasing temperature as shown in Fig. \ref{fig7}.

\subsection{Coupled ion-electron broadening}
\label{subsecIIIc}

    It has been shown in a previous section that classical two
    component MD simulations provide relevant results regarding charge
    structure and dynamics. The plasma conditions have been selected
    to emphasize screening mechanisms in order to demonstrate
    significative effects. Other results obtained for weaker coupling
    conditions lead to the same observations. These investigations
    ensure that ion plus electron field samplings are properly
    generated by the MD simulation for the plasma conditions of this
    work. Like for the EBT, field sampling is the first stage of the
    ion electron broadening theory (IEBT). In the second stage these
    fields are used for integrating the Schr\"odinger equation in order
    to synthesize lineshapes. This protocol requires a numerical cost
    much smaller than for the impurity case because 1) a number of
    samples are generated at the same time using each ion as the
    emitter, and 2) the integration time required for $C(t)$ is
    shorter when ion broadening is accounted for.
\begin{figure}[ht]
    \includegraphics[width=10cm]{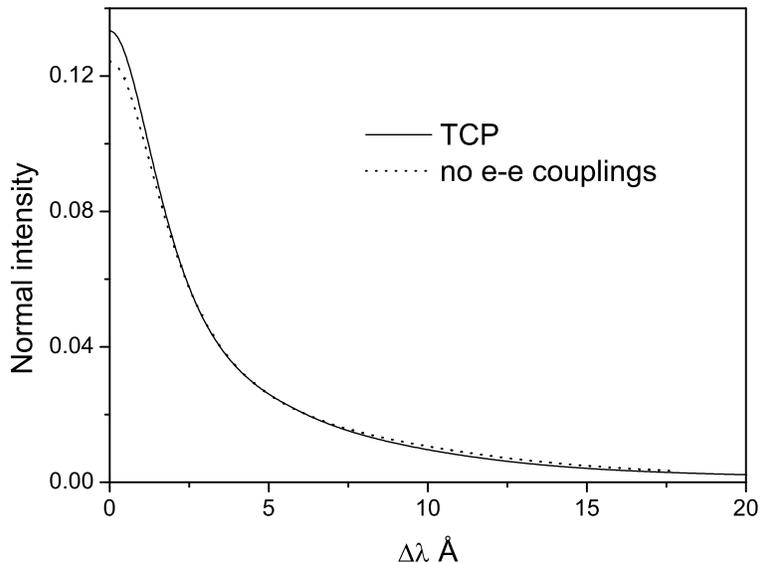}
    \caption{Helium hydrogen-like Balmer alpha line $T_{e}=50000$ K.}
    \label{fig13}
\end{figure}
    Two field data sets are used in order to look for evidences of
    ion-electron coupling mechanisms at the lineshape level. The two
    data-sets are built using TCP molecular dynamics techniques, 1)
    with all the charge-charge interactions accounted for and, 2)
    taking into account only the ion-ion interactions assuming a Debye
    screening for the ion-ion potential and free electrons moving on
    straight trajectories. The observed discrepancy in Fig.
    \ref{fig13}{} is a decrease of $6-7\%$ in the full width at half
    maximum when both the ion-electron and the electron-electron
    coupling are accounted for.

\section{Conclusion}
\label{secIV}

    Semi-classical models involving a quantum emitter perturbed by a
    stochastic classical electric field have been developed to
    calculate emitter dipole relaxation. On this basis charge
    correlation effects relevant for plasma spectroscopy have been
    explored. Such reference results from ideal numerical experiments
    are indispensable, particularly when the plasma conditions are
    such that the usual theoretical approximations appear
    questionable. The calculations reported here are for two distinct
    models, with and without ion-electron and electron-electron
    couplings. They give rise to comparisons of two sets of line
    broadening calculations. The observed residual discrepancies
    confirm that effects of charge-charge couplings on lineshapes can
    be non negligible.

    Common electron broadening theories do not account for
    ion-electron couplings. The main reason is that for most of the
    ionic lines (normal lines) electron broadening is a small effect
    with respect to ion broadening. This has not motivated the
    development of theories accounting for the complex problem of the
    quantum properties of a charged emitter undergoing a perturbation
    resulting from the electrons around the charge. However, this
    interest exists for lines whose broadening is dominated by the
    electrons. In this work charge-charge coupling effects on ionic
    lines have been investigated considering electron broadening of
    lines with ion broadening. It is worth analyzing the main results
    obtained here with the impact theory for comparison. This theory
    is founded on the hypothesis that strong collisions do not overlap
    in time. Collision operators for electron broadening have been
    derived on this basis using either linear or parabolic electron
    trajectories \cite {alexiou}. For the electrons attracted to the
    ion impurity the impact theory postulate of binary encounters is
    no longer correct as the close electron trajectories depend on the
    other close electrons via kinetic energy exchanges, particularly
    in the case of temporary orbiting trajectories. Accordingly it
    can be concluded that ion-electron couplings cannot be treated
    ignoring electron-electron couplings. In terms of the electric
    field dynamics, these couplings induce an increase of both the
    covariance and the field fluctuation rate when the emitter charge
    or the temperature increase. These effects result in a decrease of
    the width of the studied lines in comparison with the uncoupled
    case.

    Both the EBT and the IEBT have been designed to account for all
    charge-charge couplings. They are based on the numerical integration
    of the Schr\"odinger equation over a large electric field sample set
    generated by molecular dynamics techniques. For the selected lines
    and plasma conditions the fast fluctuation limit provides a simple
    theoretical expression for electron broadening in terms of the time
    integral of the field autocorrelation function. This approximation
    has been checked by comparisons with a few results obtained by full
    integration of the Schr\"odinger equation. In parallel to this work MD
    simulations and theoretical studies of the FCF for an ion impurity
    in jellium have been carried out with preliminary results reported
    in \cite{dufty-ilya}. Thus, the FCF and the fast fluctuation limit
    provide an efficient prediction of the linewidth for this simple
    impurity case. In contrast, it appears that integration of the
    Schr\"odinger equation appears easier when both ion and electron
    broadening occur together, using two component plasma MD simulations
    for the field histories. It has been shown here that such TCP
    simulations reliably yield the proper dynamic electron screening of
    the ion-ion forces.

    Finally, it can be noted that the direct integration of the
    Schr\"odinger equation for lineshapes can be rather intricate and
    expensive. Despite the apparent difficulty of this step by step
    integration process, it appears robust enough to give accurate
    results that can be seen as ideal experiments providing reference
    data appropriate to discuss or benchmark other theories.

\begin{acknowledgments}
    Support for this research has been provided in part by an European
    action integrated Picasso plan.
    The Spanish group has been financed
    by de DGICYT (project FTN2001-1827) and the Junta de Castilla y
    Le\'on (project VA009/03).
    The research of J. Dufty was supported by U. S. Department of
    Energy Grant DE-FG02ER54677 and by the University of Provence.
\end{acknowledgments}

\appendix

\section{Fast fluctuation limit}
\label{secA}

    In this Appendix an outline of the fast fluctuation limit for the
    line profile is given. The terminology ``fast fluctuation limit''
    used instead of the alternative ``impact limit'' has been
    preferred to avoid confusing it with the impact theories. The
    later relies on the calculation of an average binary collision -
    the collision operator. The impact theories generally postulate
    non overlapping strong collisions, which is not a limitation of
    the fast fluctuation limit obtained here.

\subsection{Characteristic time scales}
\label{subseca1A}

    Three characteristic time scales have to be considered. First, the
    typical correlation time of the perturbing field $\mathbf{E}(t)$,
    and in consequence, the characteristic time of $L(t)$. This time
    scale is ruled by the kinetics of the charged particles in the
    plasma. Its order of magnitude is
    \begin{equation}
    \label{EQ:ES:23}
        \tau_c \simeq \frac{r_0}{v_0}
    \end{equation}
    where $r_0$ denotes typical inter-particle distance, $v_0$ mean
    thermal velocity.

    A second time scale is fixed by the correlations of the
    dipole-moment $\mathbf{q}(t)$. The spectral width is determined by
    this lifetime, $\tau_d$, which  is, in a way, the ``unknown
    variable'' of the problem.

    Finally, a time scale is fixed by characteristic values of $L$~:
    \begin{equation}
    \label{EQ:ES:26}
        \frac{1}{\tau_H} \simeq L \simeq
        \frac{1}{\hbar}eE_0\frac{n^2a_0}{Z}
    \end{equation}
    $a_0$ is the B\"{o}hr radius, $n$ the principal quantum number of
    the upper group of states. Evolution of the dipole moment
    $\mathbf{q}(t)$ would be fixed by this frequency scale if the
    perturbing electric field were stationary.

    The relationship between $\tau_H$ and $\tau_c$ will determine the
    relevant physical phenomenon in spectral line broadening.
    $\tau_c\geq\tau_H$ is the condition for ``quasistatic'' broadening
    which is a close limit for ion broadening. Both shape and width of
    the spectral line are fixed by the statistical distribution of the
    perturbing fields, then $\tau_d\sim\tau_H$.

    If $\tau_c\ll\tau_H$ the evolution of  the dipole moment
    $\mathbf{q}(t)$ is much slower than the evolution of the
    perturbing fields. This condition fits the electron broadening
    mechanisms. In this case, perturbations are less efficient, since
    the emitter responds to a time average of the electric field
    smaller than its statistical typical value. It is the fast
    fluctuation regime. In the next section it will be seen that in
    this regime the dipole autocorrelation fits well a decreasing
    exponential whose lifetime $\tau_d\gg\tau_H >\tau_c$, is
    determined by the autocorrelation function of the perturbing
    field. It should be noted that in the present study the ratio
    $\tau_{c}/\tau_{d}$ varies from a few hundreds to a few thousands.
    Such a large ratio guarantees that the FFL is totally relevant as
    the discrepancy of the dipole autocorrelation function to an
    exponential (or equivalently the discrepancy of the lineshape to a
    Lorentzian) is non significant.

\subsection{Fast fluctuation limit}
\label{subsecA2}

    The dipole correlation function of the text can be rewritten as
    \begin{equation}
    \label{A.1}
        C(t)={\rm tr}\;\{\,{\bf q}(t)\cdot {\bf q}\rho\,\} =
            {\rm tr}_e\;\left(\,{\rm tr}_p\;\{\,e^{Lt}\rho{\bf q}\,\}
        \right) \cdot {\bf q}
    \end{equation}
    The trace has been separated into that for the emitter and plasma
    subspaces, stationarity of the Gibbs ensemble under the dynamics
    has been used, and the Liouville operator $L$ generating the
    dynamics has been introduced. It is the sum of the Liouville
    operator for the emitter in the relative coordinate system,
    $L_{e}$, the Liouville operator for the plasma including the
    center of mass (point monopole) of the emitter, $L_{p}$, and  the
    coupling between the internal emitter states and the plasma,
    $L_{i}$
    \begin{equation}
    \label{A.2}
        L = L_{e}+L_{p}+L_{i}
    \end{equation}
    Define the interaction representation generator by%
    \begin{equation}
    \label{A.3}
        e^{Lt} = e^{\left( L_{e}+L_{p}\right) t}U(t)
            = e^{\left( L_{e}+L_{p}\right) t}\;T\;
        \exp \int_{0}^{t}dt'L_{i}
        (t').
    \end{equation}
    Here $T$ is the time ordering operator with largest times to the
    left. Then
    \begin{equation}
    \label{A.4}
        C(t) \equiv {\rm tr}_{e}\;
            \left(\,
            e^{(L_{e}+L_{p})t}\,e^{\widehat{O}(t)}\,f_{e}{\bf q}
        \,\right) \cdot {\bf q}
    \end{equation}
    where  $\widehat{O}(t)$ is the average generator for the
    interaction dynamics in the atomic subspace
    \begin{equation}
    \label{A.5}
        e^{\widehat{O}(t)} = {\rm tr}_{p}\;T\;
            \exp \int_{0}^{t}dt'
        L_{i}(t')\rho f_{e}^{-1}
        = \langle
            T\;\exp \int_{0}^{t}dt'L_{i}(t')
          \rangle _{p},
          \hspace{0.3in}
          f_{e} = {\rm tr}_{p}\;\rho
    \end{equation}

    The leading terms in a cumulant expansion of this average are
    \begin{equation}
    \label{A.6}
        \widehat{O}(t)=\widehat{O}^{(1)}(t)+\widehat{O}^{(2)}(t)+..
    \end{equation}
    with
    \begin{equation}
    \label{A.7}
        \widehat{O}^{(1)}(t)=\int_{0}^{t}dt'
            \langle
            L_{i}(t')
        \rangle _{p}
    \end{equation}
    \begin{equation}
    \label{A.8}
        \widehat{O}^{(2)}(t) = \int_{0}^{t}dt'
            \int_{0}^{t'}dt^{\prime \prime }\;
        \langle
            \widetilde{L}_{i}(t')
            \widetilde{L} _{i}(t^{\prime\prime})
        \rangle _{p}
        = t\,\int_{0}^{t}dt'
        \left(1-\frac{t'}{t}\right)
        \langle
            \widetilde{L}_{i}(t)\widetilde{L}_{i}(t')
        \rangle _{p}
    \end{equation}
    where
    \begin{equation}
    \label{A.9}
        \widetilde{L}_{i}(t') =
        L_{i}(t')-{\rm tr}_{p}\;
        L_{i}(t')\rho f_{e}^{-1}
    \end{equation}

    The fast fluctuation limit corresponds to the case where the
    perturber dynamics varies rapidly on the time scale of the dipole
    correlation function $C(t).$ Since the average is performed only
    over the plasma degrees of freedom this time scale is controlled
    by the perturbers. Then the fluctuation in Eq. (\ref{A.8}) decays
    to zero for $t'=\tau \ll t$ and the factor $t'/t$
    in the integrand can be neglected
    \begin{equation}
    \label{A.10}
        \widehat{O}^{(2)}(t)\rightarrow t\int_{0}^{t}dt'
            \langle
            \widetilde{L}_{i}(t)\widetilde{L}_{i}(t')
        \rangle _{p}
    \end{equation}
    For similar reasons, higher order terms in the cumulant expansion
    are expected to be of higher order in $\tau /t$ and therefore
    negligible.

    Further simplifications occur with additional assumptions. If the
    interaction is only through a coupling of the emitter dipole to the
    plasma field then $L_{i}=-{\bf E}\cdot {\bf D}$. Also, if the
    emitter-plasma interactions are neglected in $\rho f_{e}^{-1}$ then
    \begin{equation}
    \label{A.12}
        \rho f_{e}^{-1}\rightarrow \rho _{p}
    \end{equation}
    and the fluctuations become
    \begin{eqnarray}
    \label{A.13}
        \langle L_{i}(t) \rangle _{p} & \rightarrow & 0,
        \nonumber\\
        \langle \widetilde{L}_{i}(t)\widetilde{L}_{i}(t')
        \rangle _{p} & \rightarrow &
            \frac{1}{3}\Gamma(t-t')
            {\bf D}(t){\bf\cdot D}(t')
    \end{eqnarray}
    where $\Gamma(t)$ is the electric field autocorrelation function
    for the plasma alone
    \begin{equation}
    \label{A.14}
        \Gamma (t) = \langle
            {\bf E}(t)\cdot{\bf E(0)}\rangle _{p}
    \end{equation}
    and the dipole operator in ${\bf D }(t)$ has a time dependence due
    to the emitter alone. Finally, in the case of an emitter with
    upper and lower degenerate manifolds such as the Balmer alpha line
    without fine structure this time dependence of the dipole
    operators can be neglected as well. The result quoted in the text
    is then obtained
    \begin{equation}
    \label{A.15}
        C(t) \equiv {\rm tr}_{a}\;
            \left(
            e^{L_{e}t}
            e^{t\left(\int_{0}^{\infty}dt'\,
                \Gamma (t')\,{\bf D\cdot D}\right)}
            f_{e}{\bf q}\right)\cdot{\bf q}
    \end{equation}

\pagebreak

\end{document}